\documentclass[11pt]{article}
\usepackage{amssymb}

\usepackage{graphicx}
\usepackage{amsmath}

\input{tcilatex}

\begin{document}

\begin{flushright}  Nov 24, 2000 \\
Miami TH/1/00
\end{flushright}\ 

\ 

\ 

\ 

\ 

\begin{center}
{\LARGE Schr\"{o}dinger's Cataplex} \footnote{%
A talk given at Orbis Scientiae, 17 December 1999, Fort Lauderdale, Florida.
\ To appear in the proceedings, ``Quantum Gravity, Generalized Theory of
Gravitation and Superstring Theory - Based Unification'', B. Kursunoglu and
A. Perlmutter, Eds., Plenum Press, 2000.}
\end{center}

\ 

\begin{center}
{\Large Thomas Curtright}\thinspace \footnote{%
curtright@physics.miami.edu}
\end{center}

\ 

\begin{center}
University of Miami

Department of Physics

1320 Campo Sano Drive

Coral Gables, FL 33146
\end{center}

\ 

\begin{center}
\textbf{ABSTRACT}
\end{center}

\ 

\begin{quotation}
We discuss elementary entwiners that cross-weave the variables of certain
integrable models: \ Liouville, sine-Gordon, and sinh-Gordon field theories
in two-dimensional spacetime, and their quantum mechanical reductions. \
First we define a complex time parameter that varies from one energy-shell
to another. \ Then we explain how field propagators can be simply expressed
in terms of elementary functions through the combination of an evolution in
this complex time and a duality transformation.
\end{quotation}

\ 

\noindent\textbf{IT'S COMPLEX TIME}

\ 

One hundred years ago at the close of the 19$^{th}$ century, just before
Planck's discovery of light quanta, H. M. Macdonald \cite{Macdonald}
considered the mathematical problem of determining zeroes of Bessel
functions in the complex plane. \ He was led to find the lovely integral
identity
\begin{equation*}
K_{\nu}(e^{x})\,K_{\nu}(e^{y})=\int_{-\infty}^{+\infty}dz\,S\left(
x,y,z\right) \,K_{\nu}(e^{z})\;. 
\end{equation*}
The kernel in the integral is a simple, symmetric exponential of
exponentials. 
\begin{equation*}
S\left( x,y,z\right) =\frac{1}{2}\,\exp\left( -F\left( x,y,z\right) \right)
\;,\;\;\;\;\;F\left( x,y,z\right) =\frac{1}{2}\,\left(
e^{x+y-z}+e^{x-y+z}+e^{-x+y+z}\right) \;. 
\end{equation*}
The most direct way to prove this result is through the use of the
Heine-Schl\"{a}fli identity that cogently expresses modified Bessel
functions as an integral transform\ (cf.\ \cite{Watson}, \S6.22 and
\S13.71). 
\begin{equation*}
K_{\nu}(e^{x})=K_{-\nu}(e^{x})=\frac{1}{2}\int_{-\infty}^{\infty}dX\,\exp%
\left( -e^{x}\cosh X+\nu X\right) \;. 
\end{equation*}
When this is substituted for each of the Bessel functions in the previous
bilinear, a simple change of integration variables immediately yields
Macdonald's identity. \ (For another derivation, which overlaps with many of
the standard textbook methods \cite{McLachlan, Meixner} of obtaining similar
integral relations for Mathieu functions, see \cite{Davis and Ghandour}.)

More recently in the 20$^{th}$ century, the modified Bessel functions in
Macdonald's identity have appeared in a physical context as solutions of the
Liouville quantum mechanics. \ For Liouville quantum mechanics the
Hamiltonian is $H=p^{2}+e^{2x}\;.$ \ Coordinate space energy eigenfunctions
are then solutions of 
\begin{equation*}
\left( -\frac{d^{2}}{dx^{2}}+e^{2x}\right) \psi_{E}(x)=E\,\psi_{E}(x)\;. 
\end{equation*}
For $0<E<\infty$ the bounded solutions\footnote{%
There are also unbounded solutions, $I_{\nu}$, for which Macdonald's
integral relation has a ``sister'' identity: $\theta\left( y-x\right)
\,I_{\nu}(e^{x})\,K_{\nu}(e^{y})+\theta\left( x-y\right)
\,I_{\nu}(e^{y})\,K_{\nu}(e^{x})=\int_{-\infty }^{+\infty}dz\,S\left(
x,y,z\right) \,I_{\nu}(e^{z})${}$\;$.} are 
\begin{equation*}
\psi_{E}(x)=\psi_{E}^{\ast}(x)=\frac{1}{\pi}\,\sqrt{\sinh\left( \pi\sqrt
{E}\right) }\,K_{i\sqrt{E}}(e^{x})\;. 
\end{equation*}
As indicated, these $\psi_{E}$'s are real. \ For $E=0$ there is no solution 
\cite{D'Hoker and Jackiw}. \ For other values of $E$ the wave functions are
ortho-normalized such that $\int_{-\infty}^{+\infty}dx\,\psi_{E_{1}}(x)\,%
\psi_{E_{2}}(x)=\delta(E_{1}-E_{2})$. \ The wave functions are also complete
on the appropriate space of bounded wave functions\footnote{%
A proof of completeness is given in \cite{Grosche and Steiner 87}, for
example.}, such that $\int_{0}^{\infty}dE\,\psi_{E}(x)\psi_{E}(y)=\delta%
\left( x-y\right) $.

From the reality and completeness of the Liouville wave functions for real $x
$, $y$, and $z$, it follows that another way to state Macdonald's identity%
\footnote{%
The Liouville-to-free-particle transformation kernel is obtained by taking
the limit of $F$ as $y,z\rightarrow-\infty,$ with $x$ and $X\equiv y-z$
fixed. \ This gives $F\left( x,y,z\right) \rightarrow e^{x}\cosh X$, with
the variable $X$ acting as the free particle coordinate in the
Heine-Schl\"{a}fli transform previously noted.} is 
\begin{equation*}
S\left( x,y,z\right) =\int_{0}^{\infty}dE\,K_{i\sqrt{E}}(e^{z})\;\psi
_{E}(x)\psi_{E}^{\ast}(y)\;. 
\end{equation*}
Upon comparing this expression with the standard form for the propagator as
a bilinear in wave functions, 
\begin{equation*}
G(x,y;t)=\int_{0}^{\infty}dE\,e^{-iEt}\;\psi_{E}(x)\psi_{E}^{\ast}(y)\;, 
\end{equation*}
a physical interpretation of Macdonald's identity is immediately apparent. \
Macdonald's kernel $S\left( x,y,z\right) $ is precisely the Liouville
propagator in the complex time plane, with the identification 
\begin{equation*}
e^{-iEt}=K_{i\sqrt{E}}(e^{z})\;. 
\end{equation*}
The parameters $t$ and $z$ are in direct correspondence. \ That is, the
propagator may be written as 
\begin{equation*}
G\left( x,y;t\right) \circeq\frac{1}{2}\exp\left( -\frac{1}{2}\left(
e^{x+y-z}+e^{x-y+z}+e^{-x+y+z}\right) \right) \;, 
\end{equation*}
where $\circeq$ signifies equality on a given energy shell for which $t=%
\frac{i}{E}\ln\left( K_{i\sqrt{E}}(e^{z})\right) $. \ Real $z$ corresponds
to complex $t$.

This elementary but nontrivial form for the propagator has the virtue of
having explicitly simple $x$ and $y$ coordinate dependence, without being
either pathological or tautological\footnote{%
By ``tautological form'' we mean $G\left( x,y;t\right) \circeq\delta\left(
x-y\right) \exp\left( -iw\right) $, where $w=tE$ on a given energy shell. \
This is a true statement, but by itself it clearly does not represent \emph{%
any} progress in determining the explicit properties of a system. Although
for at least one simple case it does lead to the familiar form for the
propagator. \ If it is applied as a kernel in an integral transform of a
free particle wave function, an integration by parts immediately produces
the well-known result: $\exp\left( it\partial^{2}/\partial x^{2}\right)
\,\delta\left( x-y\right) =\frac{1}{\sqrt{4\pi it}}\exp\left( i\left(
x-y\right) ^{2}/4t\right) \;.$}. \ While the time dependence is somewhat
mysteriously encoded in the variable $z$, the coordinate dependence is quite
transparent. \ In principle, such explicit coordinate dependence for \emph{%
any} propagator should greatly facilitate extracting the coordinate
dependence of the corresponding energy eigenstates. \ Recall that two
general methods for extracting such information from the propagator are
either to project onto a particular energy by Fourier transforming in the
time (that is, construct the Green function and examine the residues of its
poles), or to take the deep Euclidean time limit (and, say, isolate a
particular exponentially decaying term). \ Similar general methods should be
possible\footnote{%
In particular, another way to think of Macdonald's original identity is just
as a means of extracting the residues of the Liouville Green function poles.}
involving the variable $z$.

Some care is required, however, since the propagator interpretation of
Macdonald's identity implies that the relation between the variable $z$ and
the time $t$ is energy dependent, and not just through the combination $Et$.
\ For example, for large $\left| e^{z}\right| $, so long as $\left|
\arg\left( e^{z}\right) \right| <\frac{3}{2}\pi$, the asymptotic behavior of
the modified Bessel function is 
\begin{equation*}
K_{\nu}\left( e^{z}\right) \thicksim\sqrt{\frac{\pi}{2e^{z}}}%
e^{-e^{z}}\left\{ 1+\frac{4\nu^{2}-1}{8}e^{-z}+\frac{1}{2!}\frac{4\nu^{2}-1}{%
8}\frac{4\nu^{2}-9}{8}e^{-2z}+\cdots\right\} \;. 
\end{equation*}
This means, for deep Euclidean time $t=-iT$, $T\rightarrow\infty$, $%
e^{-iEt}\thicksim e^{-e^{z}-\frac{1}{2}z+\ln\sqrt{\frac{\pi}{2}}%
+\ln\{1+\cdots\}}$. \ That is 
\begin{equation*}
T\thicksim\frac{1}{E}\left( e^{z}+\frac{1}{2}z-\ln\sqrt{\frac{\pi}{2}}+\frac{%
1+4E}{8}e^{-z}+O\left( e^{-2z}\right) \right) \;. 
\end{equation*}
Curves in the complex $z$ plane which correspond to real time evolution
would be contours of constant modulus for $K_{i\sqrt{E}}(e^{z})$. \ If these
are open contours, the corresponding time evolution would be over an
interval, perhaps infinite. \ (If these are simple closed contours, the
corresponding time variable would be periodic, and the contours might
therefore be appropriate to describe Liouville quantum mechanics on closed
time-like curves, or perhaps at finite temperature, in the extension to
Liouville and other field theories in the following.)

Also note that this relation between $z$ and $t$ is not as strange as it
might first appear. Analytic continuation of the time variable is of course
a standard practice, but in addition, simple energy dependent (or more
generally ``state'' or ``system'' dependent) redefinitions of the time
variable are standard techniques in several areas of physics. \ For an
ancient example, in celestial mechanics\ the use of the various ``anomaly''
variables (such as the mean anomaly $l\left( t\right) =a^{-3/2}t$ where $a$
is the semi-major axis for a particular orbit \cite{Abraham and Marsden})
amounts to a somewhat trivial choice of system-dependent time. \ For a more
contemporary class of examples, general relativists routinely use
coordinates in which the new time variable depends on both the old space and
time coordinates as well as on the properties of the system in question
(e.g. black hole Kerr coordinates \cite{Wald}).

It seems odd to me that this interpretation of Macdonald's identity as a
propagator has escaped notice until now. \ Nonetheless, it would appear that
this simple interpretation has not been appreciated previously. \ For
instance, there is no mention of it among the various propagators compiled
in \cite{Grosche and Steiner 98}, although the authors of that compilation
did indeed use the previously mentioned sister identity to recast the path
integral form of the propagator into the standard sum over wave function
bilinears \cite{Grosche and Steiner 87}. \ As far as I can tell, the $%
t\leftrightarrow z$ correspondence is not realized either in the work of
Anderson, et al., \cite{Anderson, Anderson and Camporesi, Anderson et al}. \
Finally, in their recent analysis \cite{Davis and Ghandour} Davis and
Ghandour also do not make any connection to the propagator\footnote{%
Not even when they treat a special limiting case where the system under
study reduces to the simple harmonic oscillator!}. \ All in all, it would
seem that a more careful and thorough analysis of the $t\leftrightarrow z$
correspondence is warranted.

\ 

\ 

\noindent\textbf{DUALITY ENTWINES}

\ 

When this correspondence is extended to the field theory case, a more
compelling collection of ideas emerges. \ In $1+1$ field theory, a
correspondence between $z$ and complex time also exists when conjoined with
a duality transformation. \ This follows from replacing the $x$ and $y$
variables in the exponentials of Macdonald's kernel with local fields $%
\phi\left( \sigma\right) $ and $\psi\left( \sigma\right) $, integrating over
the spatial coordinate $\sigma$, and adding the elementary duality generator 
$\int\phi\partial_{\sigma}\psi d\sigma$ as follows \footnote{%
The Liouville-to-free-field transformation kernel is obtained by taking the
limit as $\psi,z\rightarrow-\infty,$ with $\phi$ and $\varphi\equiv\psi-z$
fixed. \ This gives $\frak{S}\left( \phi,\psi,z\right) \rightarrow\exp i\int
d\sigma\left( \phi\partial_{\sigma}\varphi+e^{\phi}\sinh\varphi\right) $ $,$
with the variable $\varphi$ acting as the free field. \ The corresponding
functional integral transforms discussed below become in this limit those
first used in \cite{Curtright and Ghandour} to compute certain Liouville
energy eigenfunctional matrix elements.}. 
\begin{equation*}
\frak{S}\left( \phi,\psi,z\right) \equiv\exp i\int d\sigma\left(
\phi\partial_{\sigma}\psi+\frac{1}{2}\left( e^{\phi+\psi-z}-e^{\phi-\psi
+z}-e^{-\phi+\psi+z}\right) \right) \;. 
\end{equation*}
We have also adjusted the phases in the exponential to conform to those in
the usual Schr\"{o}dinger equation, shifting $z\rightarrow z+i\pi/2$.

The kernel $\frak{S}$ ``entwines'' or ``cross-weaves'' various field theory
operators for the $\phi$ and $\psi$ fields within the Schr\"{o}dinger
wave-functional framework, as we now explain. \ We first observe that 
\begin{align*}
-i\,\frac{\delta}{\delta\phi\left( \rho\right) }\frak{S}\left( \phi
,\psi,z\right) & =\left( \partial_{\rho}\psi+\frac{1}{2}\left(
e^{\phi+\psi-z}-e^{\phi-\psi+z}+e^{-\phi+\psi+z}\right) \right) \frak{S}%
\left( \phi,\psi,z\right) \;, \\
-i\,\frac{\delta}{\delta\psi\left( \rho\right) }\frak{S}\left( \phi
,\psi,z\right) & =\left( -\partial_{\rho}\phi+\frac{1}{2}\left(
e^{\phi+\psi-z}+e^{\phi-\psi+z}-e^{-\phi+\psi+z}\right) \right) \frak{S}%
\left( \phi,\psi,z\right) \;,
\end{align*}
where we have assumed that $\frac{\delta}{\delta\psi}\left( \phi\psi\right)
=0$ at the ends of the $\sigma$ integration range. \ These results
immediately allow us to show that the kernel relates local momentum density
operators for the two fields.

By definition these momentum operators are $\mathcal{P}_{\phi }\left( \rho
\right) =-i\partial _{\rho }\phi \,\frac{\delta }{\delta \phi \left( \rho
\right) }$ and $\mathcal{P}_{\psi }\left( \rho \right) =-i\partial _{\rho
}\psi \,\frac{\delta }{\delta \psi \left( \rho \right) }$, so that 
\begin{align*}
\mathcal{P}_{\phi }\left( \rho \right) \frak{S}\left( \phi ,\psi ,z\right) &
=\partial _{\rho }\phi \,\left( \partial _{\rho }\psi +\frac{1}{2}\left(
e^{\phi +\psi -z}-e^{\phi -\psi +z}+e^{-\phi +\psi +z}\right) \right) \frak{S%
}\left( \phi ,\psi ,z\right) \;, \\
\mathcal{P}_{\psi }\left( \rho \right) \frak{S}\left( \phi ,\psi ,z\right) &
=\partial _{\rho }\psi \,\left( -\partial _{\rho }\phi +\frac{1}{2}\left(
e^{\phi +\psi -z}+e^{\phi -\psi +z}-e^{-\phi +\psi +z}\right) \right) \,%
\frak{S}\left( \phi ,\psi ,z\right) \;.
\end{align*}
Combining these and assuming that $z$ is \emph{not} also a local field, so
that $\partial _{\rho }z=0$, we obtain 
\begin{equation*}
\left( \mathcal{P}_{\phi }\left( \rho \right) +\mathcal{P}_{\psi }\left(
\rho \right) \right) \frak{S}\left( \phi ,\psi ,z\right) =\frak{S}\left(
\phi ,\psi ,z\right) \;i\partial _{\rho }F\left( \phi ,\psi ,z+i\pi
/2\right) \;.
\end{equation*}
We now integrate this last equation over $\rho $ to obtain the total, global
momentum operators for the two fields: \ $\mathbf{P}_{\phi }\mathbf{=}\int
d\rho \,\mathcal{P}_{\phi }\left( \rho \right) $, $\mathbf{P}_{\psi }\mathbf{%
=}\int d\rho \,\mathcal{P}_{\psi }\left( \rho \right) $. \ 

If we impose boundary conditions in $\rho$ such that $0=\int d\rho
\,\partial_{\rho}F\left( \phi,\psi,z+i\pi/2\right) $, then we have 
\begin{equation*}
\mathbf{P}_{\phi}\frak{S}\left( \phi,\psi,z\right) =-\mathbf{P}_{\psi }\frak{%
S}\left( \phi,\psi,z\right) \;. 
\end{equation*}
Using functional integration by parts, this implies that the two momenta are
exchanged by, or entwined with the kernel in an integral transform. \ That
is, the momenta of $\phi$ and $\psi$ wave functionals become cross-weaved
with one another when these functionals are related through the use of $%
\frak{S}\left( \phi,\psi,z\right) $ in a functional integral transform. \ \
More explicitly, let\footnote{%
Recall\ from above real $z$ corresponds to Euclidean time. \ So for
arbitrary $z$ this transformation does not necessarily preserve the
wave-functional normalizations, and hence is a similarity transformation,
but not necessarily a unitary one \cite{Curtright}. \ Nonetheless, this does
not effect the present discussion.}
\begin{equation*}
\Phi\left( \phi\right) =\int D\psi\,\frak{S}\left( \phi,\psi,z\right)
\,\Psi\left( \psi\right) \;. 
\end{equation*}
Then 
\begin{equation*}
\mathbf{P}_{\phi}\Phi\left( \phi\right) =\int D\psi\,\mathbf{P}_{\phi }\frak{%
S}\left( \phi,\psi,z\right) \,\Psi\left( \psi\right) =\int D\psi\,\frak{S}%
\left( \phi,\psi,z\right) \,\mathbf{P}_{\psi}\Psi\left( \psi\right) \;, 
\end{equation*}
where we have discarded the surface terms arising from the functional
integration by parts. \ 

This last relation is precisely what we mean by the kernel entwining or
cross-weaving the momentum operators in the Schr\"{o}dinger functional
framework \footnote{%
Hence the title of this talk. \ From Liddell-Scott-Jones Lexicon of
Classical Greek: \ $\kappa\alpha\tau\alpha\pi\lambda\varepsilon \kappa\omega$
(kataplek\^{o}) - \textit{entwine, plait. \ }Or from Herodotus \emph{%
Histories} (Loeb) [3.98.4]:
\par
\begin{quote}
Houtoi men d\^{e} t\^{o}n Ind\^{o}n phoreousi esth\^{e}ta phlo\"{i}n\^{e}n:
epean ek tou potamou phloun am\^{e}s\^{o}si kai kops\^{o}si, to entheuten
phormou tropon \textbf{kataplexantes} h\^{o}s th\^{o}r\^{e}ka endunousi.
\par
(These Indians wear clothes of bullrushes; they mow and cut these from the
river, then \textbf{having woven them crosswise} like a mat, wear them like
a breastplate.)
\end{quote}
\par
\noindent Note that kataplexantes is the plural of the active participle of
kataplek\^{o}, which we have chosen to distill for obvious reasons to a more
contemporary ``cataplex''. Of course, as scholars of classical Greek will
note, there is also: \ $\kappa\alpha\tau\alpha\pi\lambda\eta\xi$
(kata-pl\^{e}x) - \textit{stricken, struck}, usu. metaph., \textit{stricken
with amazement, astounded.} \ However, this too is an appropriate meaning
for the situation under discussion, in our opinion.
\par
After giving this talk, we learned that M. Gell-Mann had previously drawn on
the Indo-European root *plek-, from which $\pi\lambda\varepsilon\kappa\omega$
derives, to introduce the term \textit{plectics} for ``a broad
transdisciplinary subject covering aspects of simplicity and complexity as
well as the properties of complex adaptive systems''. \ (cf.
http://www.santafe.edu/sfi/People/mgm/plectics.html)}.

A perspicacious observer would notice that the momenta of $\phi$ and $\psi$
theories are entwined in precisely the same way for any choice $\frak{S}%
=\exp i\int d\sigma\left( \phi\partial_{\sigma}\psi+f\left( \phi,\psi\right)
\right) $ regardless of the form of $f\left( \phi,\psi\right) $. \ The
crucial behavior is provided solely by the elementary duality generator $%
\int d\sigma\phi\partial_{\sigma}\psi$, which by itself would interchange
spatial derivatives of the fields with their canonically conjugate variables
(realized as functional derivatives here) just as in the classical theory. \
So, cross-weaving spatial momentum operators inside the functional integral
transform is a relatively trivial task that places only minor restrictions
on the kernel. \ In a much less trivial way, the previous kernel $\frak{S}%
\left( \phi,\psi,z\right) $ also entwines with the energy operators for the $%
\phi$ and $\psi$ fields.

To demonstrate this, we need to take second functional derivatives. \
Rewrite the previous first functional derivatives as 
\begin{align*}
-i\left( \frac{\delta }{\delta \phi \left( \rho \right) }+\frac{\delta }{%
\delta \psi \left( \rho \right) }\right) \frak{S}\left( \phi ,\psi ,z\right)
& =\left\{ \partial _{\rho }\left( \psi -\phi \right) +e^{-z+\phi \left(
\rho \right) +\psi \left( \rho \right) }\right\} \frak{S}\left( \phi ,\psi
,z\right) \;, \\
-i\left( \frac{\delta }{\delta \phi \left( \rho \right) }-\frac{\delta }{%
\delta \psi \left( \rho \right) }\right) \frak{S}\left( \phi ,\psi ,z\right)
& =\left\{ \partial _{\rho }\left( \psi +\phi \right) +e^{z}\left( e^{-\phi
\left( \rho \right) +\psi \left( \rho \right) }-e^{\phi \left( \rho \right)
-\psi \left( \rho \right) }\right) \right\} \frak{S}\left( \phi ,\psi
,z\right) \;,
\end{align*}
and combine these to get the second derivative 
\begin{align*}
& \left( -i\right) ^{2}\left( \frac{\delta }{\delta \phi \left( \rho
_{1}\right) }-\frac{\delta }{\delta \psi \left( \rho _{1}\right) }\right)
\left( \frac{\delta }{\delta \phi \left( \rho _{2}\right) }+\frac{\delta }{%
\delta \psi \left( \rho _{2}\right) }\right) \frak{S}\left( \phi ,\psi
,z\right) =\left\{ 2i\partial _{\rho _{2}}\delta \left( \rho _{2}-\rho
_{1}\right) \right\} \frak{S}\left( \phi ,\psi ,z\right)  \\
& +\left\{ \partial _{\rho _{2}}\left( \psi -\phi \right) +e^{-z+\phi \left(
\rho _{2}\right) +\psi \left( \rho _{2}\right) }\right\} \left\{ \partial
_{\rho _{1}}\left( \psi +\phi \right) +e^{z}\left( e^{-\phi \left( \rho
_{1}\right) +\psi \left( \rho _{1}\right) }-e^{\phi \left( \rho _{1}\right)
-\psi \left( \rho _{1}\right) }\right) \right\} \frak{S}\left( \phi ,\psi
,z\right) \;.
\end{align*}
Taking the $1\longleftrightarrow 2$ symmetric, $\rho _{1}\rightarrow \rho
,\;\rho _{2}\rightarrow \rho $ limit of this gives 
\begin{align*}
& \left( -i\right) ^{2}\left( \frac{\delta ^{2}}{\delta \phi \left( \rho
\right) ^{2}}-\frac{\delta ^{2}}{\delta \psi \left( \rho \right) ^{2}}%
\right) \frak{S}\left( \phi ,\psi ,z\right)  \\
& \equiv \frac{1}{2}\lim_{\rho _{1},\rho _{2}\rightarrow \rho }\left(
-i\right) ^{2}\left( \frac{\delta }{\delta \phi \left( \rho _{1}\right) }-%
\frac{\delta }{\delta \psi \left( \rho _{1}\right) }\right) \left( \frac{%
\delta }{\delta \phi \left( \rho _{2}\right) }+\frac{\delta }{\delta \psi
\left( \rho _{2}\right) }\right) \frak{S}\left( \phi ,\psi ,z\right)
\;+\;\left( 1\longleftrightarrow 2\right)  \\
& =\left\{ \partial _{\rho }\left( \psi -\phi \right) +e^{-z+\phi \left(
\rho \right) +\psi \left( \rho \right) }\right\} \left\{ \partial _{\rho
}\left( \psi +\phi \right) +e^{z}\left( e^{-\phi \left( \rho \right) +\psi
\left( \rho \right) }-e^{\phi \left( \rho \right) -\psi \left( \rho \right)
}\right) \right\} \frak{S}\left( \phi ,\psi ,z\right)  \\
& =\left\{ \left( \partial _{\rho }\psi \right) ^{2}-\left( \partial _{\rho
}\phi \right) ^{2}+e^{2\psi \left( \rho \right) }-e^{2\phi \left( \rho
\right) }\right\} \frak{S}\left( \phi ,\psi ,z\right) +2\frak{S}\left( \phi
,\psi ,z\right) \;\partial _{\rho }F\left( \phi ,\psi ,z\right) \;.
\end{align*}
Now the local energy density operators for the $\phi $ and $\psi $ fields
are of the same form for either: 
\begin{equation*}
\mathcal{H}_{\phi }\left( \rho \right) =-\frac{1}{2}\frac{\delta ^{2}}{%
\delta \phi \left( \rho \right) ^{2}}+\frac{1}{2}\left( \partial _{\rho
}\phi \right) ^{2}+\frac{1}{2}e^{2\phi \left( \rho \right) }\;,\;\;\;\;\;%
\mathcal{H}_{\psi }\left( \rho \right) =-\frac{1}{2}\frac{\delta ^{2}}{%
\delta \psi \left( \rho \right) ^{2}}+\frac{1}{2}\left( \partial _{\rho
}\psi \right) ^{2}+\frac{1}{2}e^{2\psi \left( \rho \right) }\;.
\end{equation*}
In view of these, the previous second derivative relation is\ 
\begin{equation*}
\mathcal{H}_{\phi }\left( \rho \right) \frak{S}\left( \phi ,\psi ,z\right) =%
\mathcal{H}_{\psi }\left( \rho \right) \frak{S}\left( \phi ,\psi ,z\right) +%
\frak{S}\left( \phi ,\psi ,z\right) \;\partial _{\rho }F\left( \phi ,\psi
,z\right) \;.
\end{equation*}
If we once again integrate over $\rho $ to obtain the total energy operators
for either field, $\mathbf{H}_{\phi }=\int d\rho \,\mathcal{H}_{\phi }\left(
\rho \right) $ and $\mathbf{H}_{\psi }=\int d\rho \,\mathcal{H}_{\psi
}\left( \rho \right) $\ , and if we again impose boundary conditions in $%
\rho $ such that $0=\int d\rho \,\partial _{\rho }F\left( \phi ,\psi
,z\right) $, we finally obtain 
\begin{equation*}
\mathbf{H}_{\phi }\frak{S}\left( \phi ,\psi ,z\right) =\mathbf{H}_{\psi }%
\frak{S}\left( \phi ,\psi ,z\right) \;.
\end{equation*}
Acting on wave functionals, this leads to 
\begin{equation*}
\mathbf{H}_{\phi }\Phi \left( \phi \right) =\int D\psi \,\mathbf{H}_{\phi }%
\frak{S}\left( \phi ,\psi ,z\right) \,\Psi \left( \psi \right) =\int D\psi \,%
\frak{S}\left( \phi ,\psi ,z\right) \,\mathbf{H}_{\psi }\Psi \left( \psi
\right) \;.
\end{equation*}
The energy operators are therefore also cross-woven by the kernel $\frak{S}%
\left( \phi ,\psi ,z\right) $.

All this entwined structure is present in other models besides the
Liouville. \ For example, the sinh-Gordon and sine-Gordon theories in $1+1$
dimensions also have simple entwining kernels\footnote{%
I obtained the kernels for the sine-Gordon and sinh-Gordon theories in the
1980's, after having spent a few years working on the quantization of the
classical B\"{a}cklund transformation connecting Liouville and free fields,
initially in an operator framework \cite{Braaten et al.} and subsequently
using functional methods in collaboration with Ghassan Ghandour and my
student Thomas McCarty. \ The latter work was not published until 1991 \cite
{Curtright and Ghandour, McCarty}.} explicitly given by exponentials of
exponentials. \ This is not surprising. \ These models are well-known \cite
{McLaughlin and Scott, Shadwick} to be the only ones (besides quadratic
Hamiltonians) involving one-component fields which can be reduced to
first-order differential equations involving two such fields whose
consistency requires the same second-order equations for either field
separately (i.e. auto-B\"{a}cklund transformations, as discussed in \cite
{Rogers and Shadwick}). \ In that purely classical context, the parameter $z$
above is known as a ``B\"{a}cklund parameter''. \ Exponentiation of the
corresponding classical generators of the first-order equations, to arrive
at quantum theories, follows from Dirac's correspondence rule \cite{Dirac}.
\ The fact that the naive correspondence works exactly, without the need for
local quantum corrections, for the Liouville, sinh-Gordon, and sine-Gordon
theories, is in our view the essence of the integrability of these models%
\footnote{%
Davis and Ghandour \cite{Davis and Ghandour} have effectively shown that
there are no other models involving one-component fields for which the
classical generators can be used in Dirac's correspondence to obtain valid
kernels. \ If more components are allowed, however, there are many more
models such that the classical generators provide entwining kernels. \ The $%
\sigma$ model is one such notable example \cite{Curtright and Zachos}.}. \
However, it should be stressed that the exact propagator for \emph{any}
theory provides an exact entwining kernel\footnote{%
We leave it as a straightforward exercise for the student to show that the
usual propagators for the linear potential and the harmonic oscillator are
indeed entwiners in the above sense, and for these simple QM examples the
correspondence between $t$ and $z$ is in fact energy \emph{in}dependent.},
even if that propagator has extensive quantum corrections, and even if those
corrections are non-local. \ This point has been noted previously \cite
{Curtright and Ghandour} with somewhat different emphasis.

Let us summarize the results for exponential potentials. \ The general form
of the kernel is: 
\begin{equation*}
\frak{S}\left( \phi,\psi,z\right) \equiv\exp i\int d\sigma\,\frak{F}\left(
\phi\left( \sigma\right) ,\psi\left( \sigma\right) ,z\right) 
\end{equation*}
\noindent The explicit forms of the generators, the local energy densities,
and their effects on the kernel are in the following table.$\smallskip$

\begin{center}
$
\begin{array}{cc}
\begin{array}{c}
\text{\textsf{Liouville}} \\ 
\\ 
\\ 
\\ 
\\ 
\smallskip 
\end{array}
& 
\begin{array}{c}
\mathcal{H}_{\phi }=-\frac{1}{2}\frac{\delta ^{2}}{\delta \phi \left( \rho
\right) ^{2}}+\frac{1}{2}\left( \partial _{\rho }\phi \right) ^{2}+\frac{1}{2%
}e^{2\phi \left( \rho \right) } \\ 
\\ 
\frak{F}\equiv \phi \partial _{\sigma }\psi +\frac{1}{2}\left( e^{-z+\phi
+\psi }-e^{z-\phi +\psi }-e^{z+\phi -\psi }\right)  \\ 
\\ 
\mathcal{H}_{\phi }\frak{S}=\mathcal{H}_{\psi }\frak{S}+\frac{1}{2}\frak{S\;}%
\partial _{\rho }\left\{ e^{-z+\phi +\psi }+e^{z-\phi +\psi }+e^{z+\phi
-\psi }\right\}  \\ 
\smallskip 
\end{array}
\\ 
\begin{array}{c}
\text{\textsf{sinh-Gordon}} \\ 
\\ 
\\ 
\\ 
\\ 
\smallskip 
\end{array}
& 
\begin{array}{c}
\mathcal{H}_{\phi }=-\frac{1}{2}\frac{\delta ^{2}}{\delta \phi \left( \rho
\right) ^{2}}+\frac{1}{2}\left( \partial _{\rho }\phi \right) ^{2}+\cosh
2\phi \left( \rho \right)  \\ 
\\ 
\frak{F}\equiv \phi \partial _{\sigma }\psi +e^{-z}\cosh \left( \phi +\psi
\right) -e^{+z}\cosh \left( \phi -\psi \right)  \\ 
\\ 
\mathcal{H}_{\phi }\frak{S}=\mathcal{H}_{\psi }\frak{S}+\frak{S\;}\partial
_{\rho }\left\{ e^{-z}\cosh \left( \phi +\psi \right) +e^{+z}\cosh \left(
\phi -\psi \right) \right\}  \\ 
\smallskip 
\end{array}
\\ 
\begin{array}{c}
\text{\textsf{sine-Gordon}} \\ 
\\ 
\\ 
\\ 
\\ 
\smallskip 
\end{array}
& 
\begin{array}{c}
\mathcal{H}_{\phi }=-\frac{1}{2}\frac{\delta ^{2}}{\delta \phi \left( \rho
\right) ^{2}}+\frac{1}{2}\left( \partial _{\rho }\phi \right) ^{2}-\cos
2\phi \left( \rho \right)  \\ 
\\ 
\frak{F}\equiv \phi \partial _{\sigma }\psi -e^{-z}\cos \left( \phi +\psi
\right) +e^{+z}\cos \left( \phi -\psi \right)  \\ 
\\ 
\mathcal{H}_{\phi }\frak{S}=\mathcal{H}_{\psi }\frak{S}+\frak{S\;}\partial
_{\rho }\left\{ -e^{-z}\cos \left( \phi +\psi \right) -e^{+z}\cos \left(
\phi -\psi \right) \right\}  \\ 
\smallskip 
\end{array}
\end{array}
$
\end{center}

\noindent The above generators for the sinh-Gordon and sine-Gordon theories
have been used in a classical context for a long time\footnote{%
The Liouville generator follows from the sinh-Gordon generator as a
contraction: shift $\phi\rightarrow\phi+w$, $\psi\rightarrow\psi+w,$ $%
z\rightarrow z+w$, rescale $\sigma\rightarrow e^{-w}\sigma$, and take $%
w\rightarrow\infty$.}. \ They appear in textbooks as the generators of
(auto) B\"{a}cklund transformations \cite{Rogers and Shadwick, Whitham},
where for the sine-Gordon case their functional derivatives are most often
employed to generate classical ($N+1$)-soliton solutions starting from $N$%
-soliton solutions, with $\phi=0$ as the trivial $N=0$ soliton \cite{Seeger
et al., Lamb}. \ In that classical situation the B\"{a}cklund parameter $z$
is related to the rapidity of the soliton's center of mass.

Now let us reconsider those total divergence terms that are produced by
entwining the densities with $\frak{S}\left( \phi ,\psi ,z\right) $ and show
that they are just the usual conformal improvements for the energy-momentum\
tensor. \ We find in all three cases: 
\begin{equation*}
\left( \mathcal{H}_{\phi }-\partial _{\rho }^{2}\phi \left( \rho \right) +%
\mathcal{P}_{\phi }-\partial _{\rho }\left( -i\frac{\delta }{\delta \phi
\left( \rho \right) }\right) \right) \frak{S}=\left( \mathcal{H}_{\psi
}+\partial _{\rho }\left( -i\frac{\delta }{\delta \psi \left( \rho \right) }%
\right) -\mathcal{P}_{\psi }-\partial _{\rho }^{2}\psi \left( \rho \right)
\right) \frak{S}
\end{equation*}
\begin{equation*}
\left( \mathcal{H}_{\phi }-\partial _{\rho }^{2}\phi \left( \rho \right) -%
\mathcal{P}_{\phi }+\partial _{\rho }\left( -i\frac{\delta }{\delta \phi
\left( \rho \right) }\right) \right) \frak{S}=\left( \mathcal{H}_{\psi
}+\partial _{\rho }\left( -i\frac{\delta }{\delta \psi \left( \rho \right) }%
\right) +\mathcal{P}_{\psi }+\partial _{\rho }^{2}\psi \left( \rho \right)
\right) \frak{S}
\end{equation*}
The pattern clearly shows that the densities for the two fields undergo dual
improvements. 
\begin{equation*}
\mathcal{H}_{\phi }\rightarrow \mathcal{H}_{\phi }-\partial _{\rho }^{2}\phi
\left( \rho \right) \;,\;\;\;\;\;\mathcal{P}_{\phi }\rightarrow \mathcal{P}%
_{\phi }-\partial _{\rho }\left( -i\frac{\delta }{\delta \phi \left( \rho
\right) }\right) \;,
\end{equation*}
\begin{equation*}
\mathcal{H}_{\psi }\rightarrow \mathcal{H}_{\psi }+\partial _{\rho }\left( -i%
\frac{\delta }{\delta \psi \left( \rho \right) }\right) \;,\;\;\;\;\;%
\mathcal{P}_{\psi }\rightarrow \mathcal{P}_{\psi }+\partial _{\rho }^{2}\psi
\left( \rho \right) \;.
\end{equation*}
Expressing this covariantly for classical densities, using $T_{\mu \nu }$
for the conventional unmodified energy-momentum tensor and $\theta _{\mu \nu
}$ for the conformally improved one, we would have the on-shell relations 
\begin{align*}
\theta _{\mu \nu }\left( \phi \right) & =T_{\mu \nu }\left( \phi \right)
+\left( g_{\mu \nu }\square -\partial _{\mu }\partial _{\nu }\right) \phi
=T_{\mu \nu }\left( \phi \right) +\varepsilon _{\mu \alpha }\varepsilon
_{\nu \beta }\partial ^{\alpha }\partial ^{\beta }\phi  \\
\theta _{\mu \nu }\left( \psi \right) & =T_{\mu \nu }\left( \psi \right) +%
\frac{1}{2}\left( \varepsilon _{\mu \alpha }\partial ^{\alpha }\partial
_{\nu }+\varepsilon _{\nu \alpha }\partial ^{\alpha }\partial _{\mu }\right)
\psi 
\end{align*}
Note that $\varepsilon _{0\alpha }\partial ^{\alpha }f=\partial _{\sigma }f$
and $\varepsilon _{1\alpha }\partial ^{\alpha }f=\partial _{\tau }f$ where $%
\sigma $ and $\tau $ are the space and time coordinates in $1+1$ dimensions.

So, with these local modifications, the energy and momentum densities are
entwined without left-over total derivatives. \ In the case of the Liouville
theory, at least, this means that the full Virasoro algebra entwines with $%
\frak{S}$. \ For an earlier quantum mechanical example of this situation,
see \cite{Curtright}.

\ 

\ 

\noindent\textbf{WISHFUL\ THOUGHTS}

\ 

Perhaps other models can be found to have simple propagators using the
approach discussed here. \ An interesting case would be the nonlinear $\sigma
$ model, and its supersymmetric siblings, which can be entwined with a dual $%
\sigma$ model at the expense of deforming the field manifold and introducing
torsion \cite{Curtright and Zachos, Curtright and Uematsu and Zachos}. \ It
is not yet known how to incorporate the parameter $z$, and by correspondence
the time, into this transformation.

Perhaps this approach to propagators is also useful when the $(\tau
,\sigma)=(\zeta^{0},\zeta^{1})$ manifold is not intrinsically flat. \ For
example, classical relations \emph{at fixed time} between the Liouville
field $\phi$ and a ``free'' field $\psi$ have been discussed before \cite
{Preitschopf and Thorn 90} (also see \cite{D'Hoker 91}). These fields
satisfy 
\begin{equation*}
D^{\mu}D_{\mu}\phi={\frac{1}{2g}}~R-{\frac{4m^{2}}{g}}~e^{2g\phi },\ \ \ \ \
\ D^{\mu}(\partial_{\mu}\psi-{\frac{1}{g}}~\omega_{\mu})=0, 
\end{equation*}
where zweibein $e_{\mu}^{a}$, connection $\omega_{\mu}=\eta_{ab}e_{\mu}^{a}%
\epsilon^{\nu\lambda}\partial_{\nu}e_{\lambda}^{b}$, and scalar curvature$\
R=-2\epsilon^{\mu\nu}\partial_{\mu}\omega_{\nu}$ are given functions that
depend on $(\zeta^{0},\zeta^{1})$. \ Canonical equivalence of the $\phi$ and 
$\psi$ fields in this curved surface situation again follows from a
generating function, which here depends explicitly on $\zeta^{0}$ even
before evolution is included. 
\begin{equation*}
F[\phi,\psi;\zeta^{0}]=\int d\zeta^{1}\left( \phi\,\left( \partial_{1}\psi-{%
\frac{1}{g}}~\omega_{1}(\zeta)\right) -\frac{2m}{g^{2}}\,e^{g\phi
}\,e_{1}^{a}(\zeta)V_{a}(\psi)\right) \ . 
\end{equation*}
The tangent space vector $V$ is given by $(V_{0},V_{1})=\left( \cosh
(g\psi),\sinh(g\psi)\right) $.

To generalize these decade-old results and find the analogue of MacDonald's
century-old propagator, we seek a generating function that yields
exponential potentials for both fields, and that allows for evolution in $%
\zeta^{0}$ through an explicit $z$ parameter. \ Indeed, such a
generalization is needed to make contact with other studies of propagators
and correlation functions for the Liouville and sine/sinh-Gordon models,
since these other studies almost invariably consider the underlying
spacetime to be a sphere. \ For example, see \cite{Goulian and Li 91, Dorn
and Otto, Zamolodchikovs 95, Lukyanov and Zamolodchikov, Fateev et al,
O'Raifeartaigh et al 98}. \ It remains to entwine all these investigations.

\ 

\ 

\noindent\textbf{Acknowledgements}

\ 

I thank Cosmas Zachos and Ghassan Ghandour for informative exchanges. \ This
work was supported in part by NSF Award 0073390.

\end{document}